\documentclass[3p,times]{elsarticle}

\usepackage{ecrc}

\usepackage{epstopdf}

\volume{00}

\firstpage{1}

\journalname{Nuclear Physics A}

\runauth{W. Florkowski et al.}

\jid{nupha}

\jnltitlelogo{Nuclear Physics A}

\usepackage{graphicx}
\usepackage{amsmath,amssymb}

\begin{document}

\begin{frontmatter}

\title{Thermalization of anisotropic quark-gluon plasma produced by decays of color flux tubes}

\author[1]{Wojciech Florkowski}
\author[2]{Radoslaw Ryblewski}

\address[1]{The H. Niewodnicza\'nski Institute of Nuclear Physics, Polish Academy of Sciences,
PL-31342 Krak\'ow, Poland \\ Institute of Physics, Jan Kochanowski University, PL-25406~Kielce, Poland}
\address[2]{The H. Niewodnicza\'nski Institute of Nuclear Physics, Polish Academy of Sciences,
PL-31342 Krak\'ow, Poland \\ Department of Physics, Kent State University, Kent, OH 44242 United States}

\begin{abstract}
Kinetic equations are used to study thermalization of the anisotropic quark-gluon plasma produced by decays of color flux tubes possibly created at the very early stages of relativistic heavy-ion collisions. The decay rates of the initial color tubes are given by the Schwinger formula, while the collision terms are taken in the relaxation-time approximation. By connecting the relaxation time with viscosity we analyze production and thermalization processes in the plasma characterized by different values of the ratio of the shear viscosity to entropy density.
\end{abstract}

\begin{keyword}
relativistic heavy-ion collisions, quark-gluon plasma, Boltzmann-Vlasov equation, anisotropic dynamics
\end{keyword}

\end{frontmatter}

\section{Introduction}
\label{intro}
In this work we study thermalization of the anisotropic quark-gluon plasma produced by decays of color flux tubes which may be created at the very early stages of relativistic heavy-ion collisions \cite{Casher:1978wy,Glendenning:1983qq,
Bialas:1984wv,Gyulassy:1986jq,Gatoff:1987uf}. Our approach is based on the kinetic theory \cite{Elze:1986qd,Elze:1986hq,Bialas:1987en,
Dyrek:1988eb,Banerjee:1989by} where decay rates of the initial color fields are given by the Schwinger formula and appear as the source terms in the kinetic equations. The produced quarks and gluons interact with the mean color field and collide with each other. The collisions are described by the collision terms treated in the relaxation-time approximation \cite{Bhatnagar:1954zz,Baym:1984np,Baym:1985tna}. 

By connecting the relaxation time with the system's shear viscosity, we are able to study production and thermalization processes in the quark-gluon plasma characterized by different values of the ratio of the shear viscosity to entropy density, ${\bar \eta} = \eta/\sigma$. 
We note that a more detailed presentation of our framework has been recently published in \cite{Ryblewski:2013eja}. We also note that the kinetic simulations with the fixed ${\bar \eta}$ have been discussed independently in \cite{Ruggieri:2013bda,Ruggieri:2013ova}. 

\begin{figure}
\begin{center}
\includegraphics*[width=6.cm]{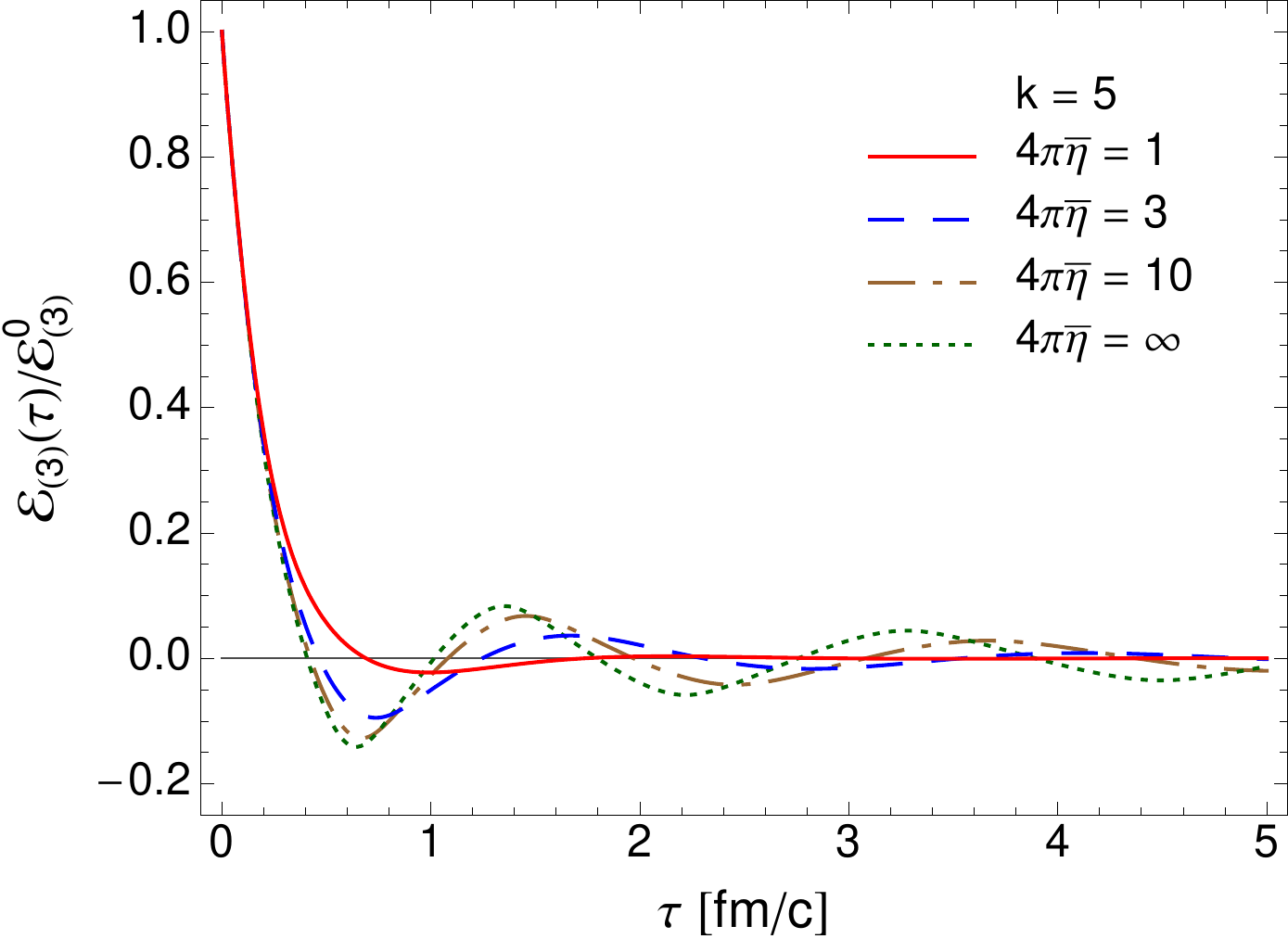}
\quad
\includegraphics*[width=6.cm]{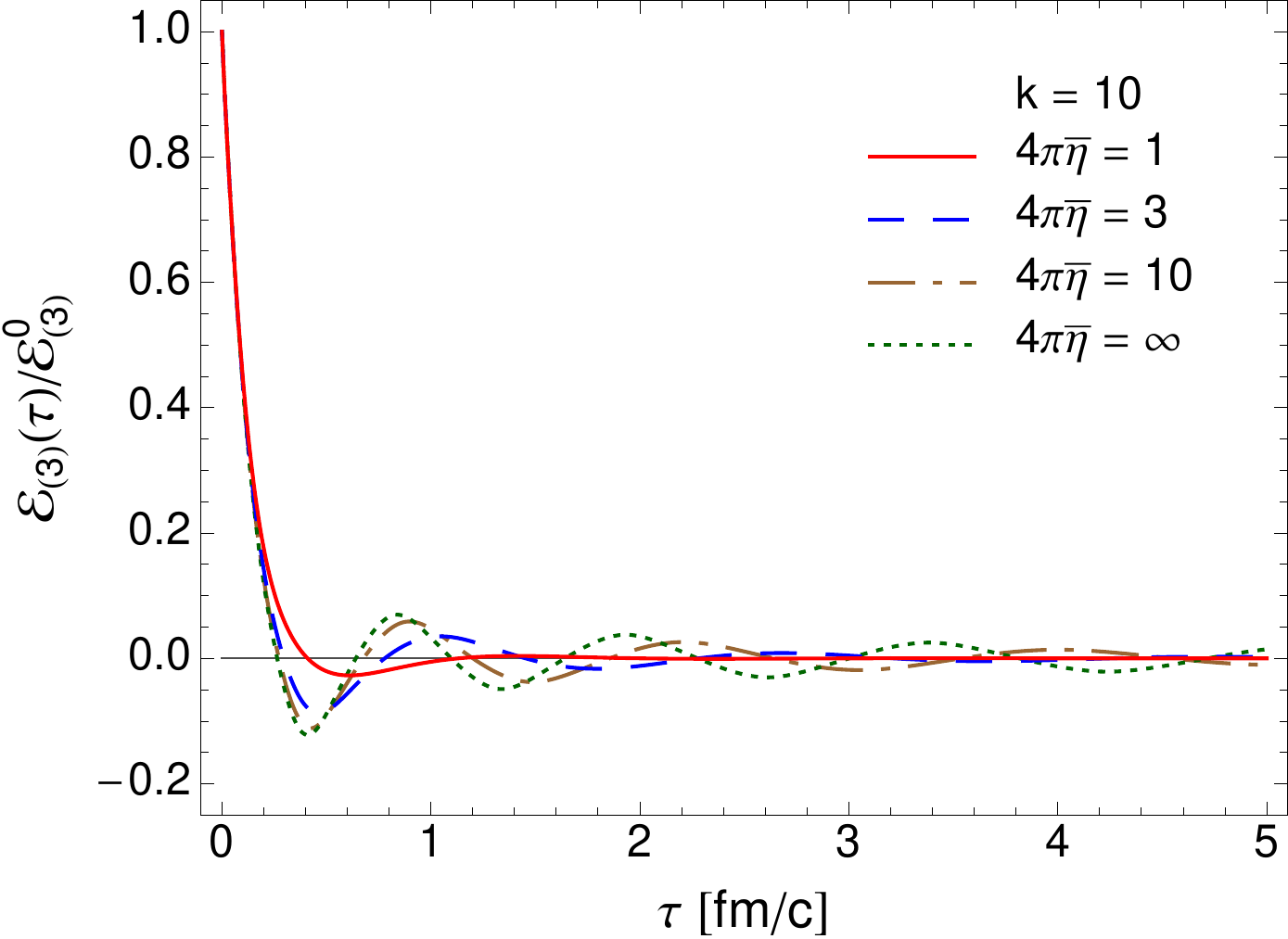}
\caption{(Color online) Time dependence of the chromoelectric field for different values of the viscosity in the case $k=5$ (left) and $k=10$ (right). The values of the field are normalized to its initial value.}
\label{fig:fields}
\end{center}
\end{figure}

\section{Color-flux-tube model}

In the color-flux tube model, the eight gluon fields are decomposed into two neutral and three charged gluon fields~\cite{Casher:1978wy}. The neutral fields ${\bf F}_{}^{\mu \nu } $ are treated classically, while the charged fields are treated as massless particles. The neutral components represent the longitudinal chromoelectric fields, ${\mbox{\boldmath $\cal E$}} = {\bf F}^{30} = (F^{30}_{(3)},F^{30}_{(8)})$. Such a field configuration has common features with the glasma phase \cite{Lappi:2006fp}, however, the chromomagnetic fields are absent in our approach.

The kinetic equations have the form \cite{Bialas:1987en,Dyrek:1988eb,Banerjee:1989by,Ryblewski:2013eja}
\begin{equation}
\left( p^\mu \partial_\mu + g{\mbox{\boldmath $\epsilon$}}_i \cdot 
{\bf F}_{}^{\mu \nu } p_\nu \partial_\mu^p\right) G_{if}(x,p) = \frac{dN_{if}}{d\Gamma_{\rm inv} }+ C_{if},\label{k1}
\end{equation}
\begin{equation}
\left( p^\mu \partial_\mu - g{\mbox{\boldmath $\epsilon$}}_i \cdot 
{\bf F}_{}^{\mu \nu } p_\nu \partial_\mu^p\right) \bar{G}_{if}(x,p) = \frac{dN_{if}}{d\Gamma_{\rm inv} } + \bar{C}_{if},\label{k2}
\end{equation}
\begin{equation}
\left( p^\mu \partial _\mu + g{\mbox{\boldmath $\eta$}}_{ij} \cdot 
{\bf F}_{}^{\mu \nu } p_\nu \partial_\mu^p\right) \widetilde{{G}}_{ij}(x,p) = 
\frac{d\widetilde{N}_{ij}}{d\Gamma_{\rm inv} }+ \tilde{C}_{ij}, \label{k3}
\end{equation}
where $G_{if}(x,p)\ $, $\bar{G}_{if}(x,p)$ and $\widetilde{{G}}_{ij}(x,p)$ are phase-space densities of quarks, antiquarks, and  gluons with color charges
\begin{equation}
\mbox{\boldmath $\epsilon$}_{1} = \frac{1}{2}\left(\! 1,\sqrt{\frac{1}{3}}\right), \quad 
\mbox{\boldmath $\epsilon$}_{2} = \frac{1}{2}\left(\! -1,\sqrt{\frac{1}{3}}\right), \quad
\mbox{\boldmath $\epsilon$}_{3} = \left(\! 0,-\sqrt{\frac{1}{3}}\right), \quad
{\mbox{\boldmath $\eta$}}_{ij}={\mbox{\boldmath $\epsilon$}}_{i}-{%
\mbox{\boldmath $\epsilon$}}_{j}. \label{charges}
\end{equation}
On the right-hand sides of Eqs.~(\ref{k1})--(\ref{k3}) the terms $dN/d\Gamma_{\rm inv}$ describe the Schwinger tunneling processes, and $C$'s are the collision terms.

The Gauss law applied to a color flux tube yields ${\mbox{\boldmath $\cal E$}} {\cal A} = k g {\bf q}$, where ${\cal A} = \pi r^2$ denotes the area of the transverse cross section of the tube, $k$ is the number of color charges at the end of the tube, and $g{\bf q} = g (q_{(3)},q_{(8)})$ is the color charge of a quark or a gluon,
see Eq.~(\ref{charges}). The Gauss law can be rewritten also as ${\mbox{\boldmath $\cal E$}}= k {\bf q} \sqrt{6 \sigma_q/(\pi r^2)} $, where $\sigma_q=1$~GeV/fm is the quark string tension of an elementary tube. This equation determines the value of the initial chromoelectric field spanned by the two receding nuclei. We use $\pi r^2$ = 1 fm$^2$ and do the calculations for $k=5$ and $k=10$, since the effective temperature of the produced plasma reaches in these two cases the values expected at RHIC and the LHC, $T_{\rm max} \sim 300-500$ MeV \cite{Ryblewski:2013eja}.

\begin{figure}
\begin{center}
\includegraphics*[width=6.cm]{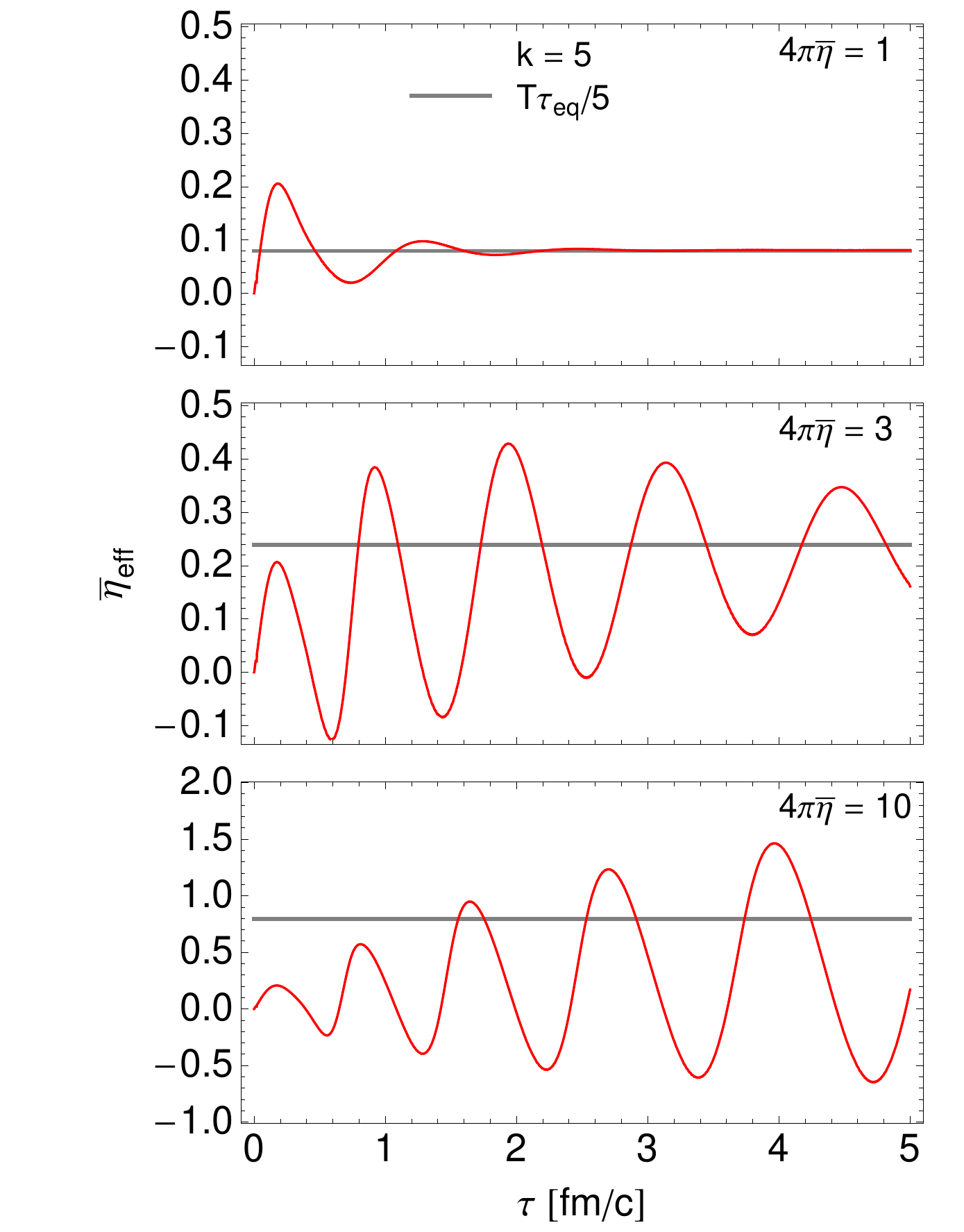}
\quad
\includegraphics*[width=6.cm]{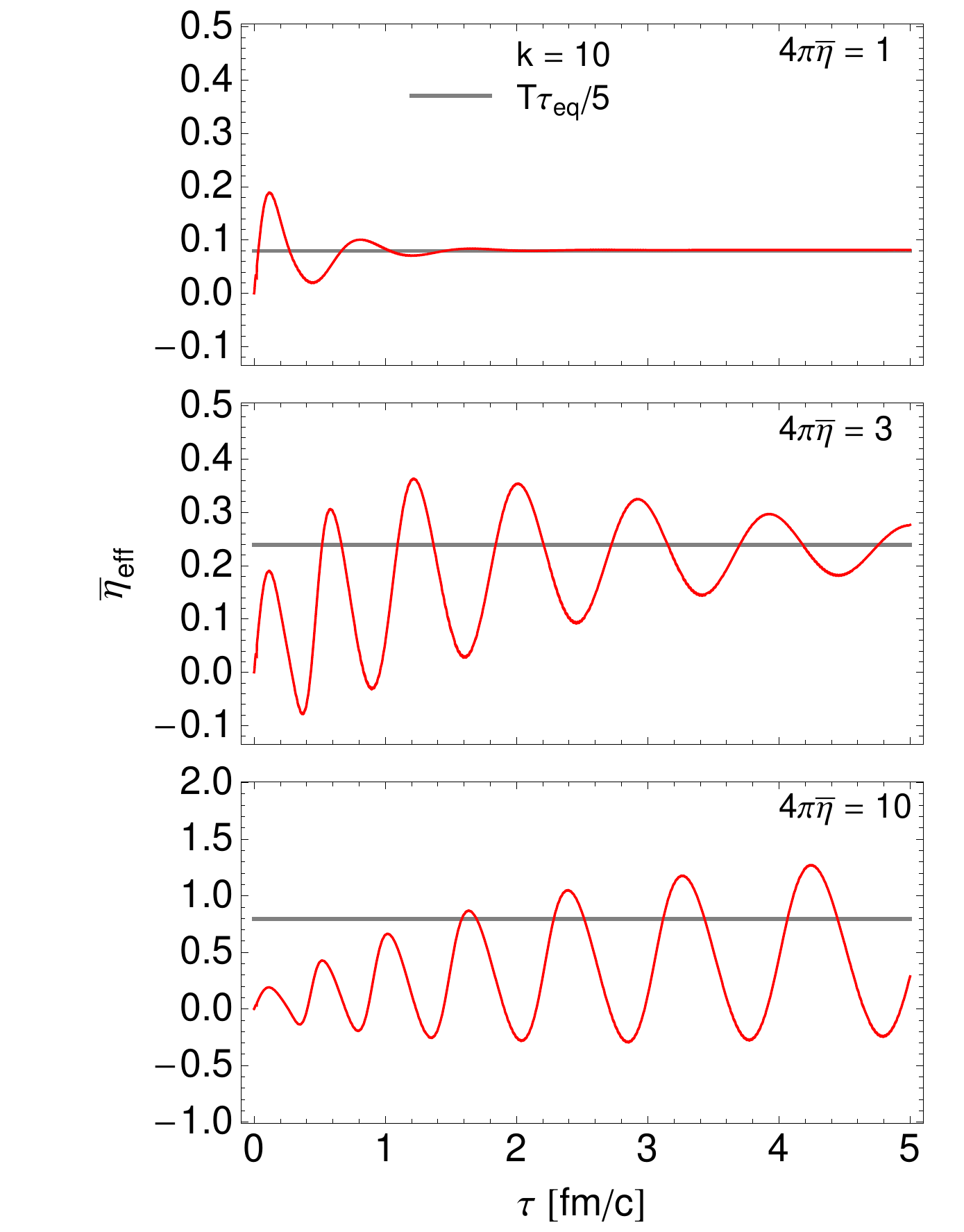}
\caption{(Color online) Time dependence of the effective viscosity ${\bar \eta}_{\rm eff}$ for different values of the parameter $4\pi{\bar \eta}$ in the case $k=5$ (left) and $k=10$ (right).}
\label{fig:visc}
\end{center}
\end{figure}

The relaxation time appearing in the collision terms $C$ has the form $\tau_{\rm eq} = 5 {\bar \eta}/T$ \cite{Anderson:1974,Czyz:1986mr,Dyrek:1986vv,
Romatschke:2011qp,Florkowski:2013lza,
Florkowski:2013lya}. 
Here ${\bar \eta}$ is the ratio of the viscosity to the entropy density which is treated as a constant in our approach. We consider three values of this ratio: ${\bar \eta} = 1/(4\pi)\,,\, 3/(4\pi)\,,  10/(4\pi)$. The first two numbers determine the viscosity range extracted from the recent hydrodynamic analyses of relativistic heavy-ion collisions studied at RHIC and the LHC. The last value is on the order expected by leading log perturbative results extrapolated to RHIC and the LHC energies. We note that the value ${\bar \eta} = 1/(4\pi)$ corresponds to a lower (KSS) bound for the viscosity found in \cite{Kovtun:2004de}.

\section{Results}

Figure~\ref{fig:fields} shows the time dependence of the chromoelectric field normalized to its initial value, calculated in the cases $k=5$ (left) and $k=10$ (right). Different curves describe our results obtained for different values of the viscosity. In the case $4\pi{\bar \eta}=\infty$ and $k=5$ we reproduce the result obtained originally in Ref.~\cite{Bialas:1987en}. As the viscosity of the system decreases, the collisions between particles become more frequent, the system becomes more dissipative, and the oscillations of the field become more and more damped. For $4\pi{\bar \eta}=1$ the oscillations practically disappear for both $k=5$ and $k=10$ (solid red lines in the left and right parts of Fig.~\ref{fig:fields}). The knowledge of the time dependence of the chromoelectric field allows us to calculate all other interesting quantities, such as the energy density, the longitudinal and transverse pressures, and the effective temperature $T$ \cite{Ryblewski:2013eja}. We note that the latter is a measure of the energy density and only if the system approaches equilibrium $T$ may be interpreted as the genuine thermodynamic quantity.

To characterize the system's behavior at later times we assume that the plasma may be characterized by the first-order viscous-hydrodynamics equations. In the case of one-dimensional boost-invariant expansion, the ratio of the shear viscosity to entropy density is  connected with the system's effective temperature $T(\tau)$  through the formula \cite{Florkowski:2013lza,Florkowski:2013lya}
\begin{eqnarray}
\frac{dT}{d\tau}+\frac{T}{3\tau}=\frac{4 {\bar \eta}_{\rm eff}}{9 \tau^2}.
\label{eta-fo}
\end{eqnarray}
We use our numerical results for $T(\tau)$ and substitute them into the left-hand side of Eq.~(\ref{eta-fo}) in order to calculate the effective viscosity ${\bar \eta}_{\rm eff}$. The functions ${\bar \eta}_{\rm eff}(\tau)$ obtained for different values of the parameters $k$ and $4\pi{\bar \eta}$ are shown in Fig.~\ref{fig:visc}. 

The two upper parts of Fig.~\ref{fig:visc} show that in the minimum viscosity case the effective viscosity of the system starts to agree very well with the viscosity parameter ${\bar \eta}$ after 1--2 fm/c, for both $k=5$ and $k=10$. Consequently, for $\tau >$ 1~fm/c our complicated system of fields and particles is very well described by the first-order viscous hydrodynamics. On the other hand, for larger values of the viscosity, for instance, in the cases $4\pi{\bar \eta}=3$ and $4\pi{\bar \eta}=10$ the effective viscosity ${\bar \eta}_{\rm eff}$ differs from the value of ${\bar \eta}$. In these cases the collisions in the plasma are inefficient to damp down the plasma oscillations. The presence of such oscillations brings in differences between the kinetic and viscous-hydrodynamics descriptions, and indicates that the viscous-hydrodynamics description after 1--2 fm/c is not completely satisfactory if $4\pi{\bar \eta} \geq 3$.

\begin{figure}
\begin{center}
\includegraphics*[width=6.cm]{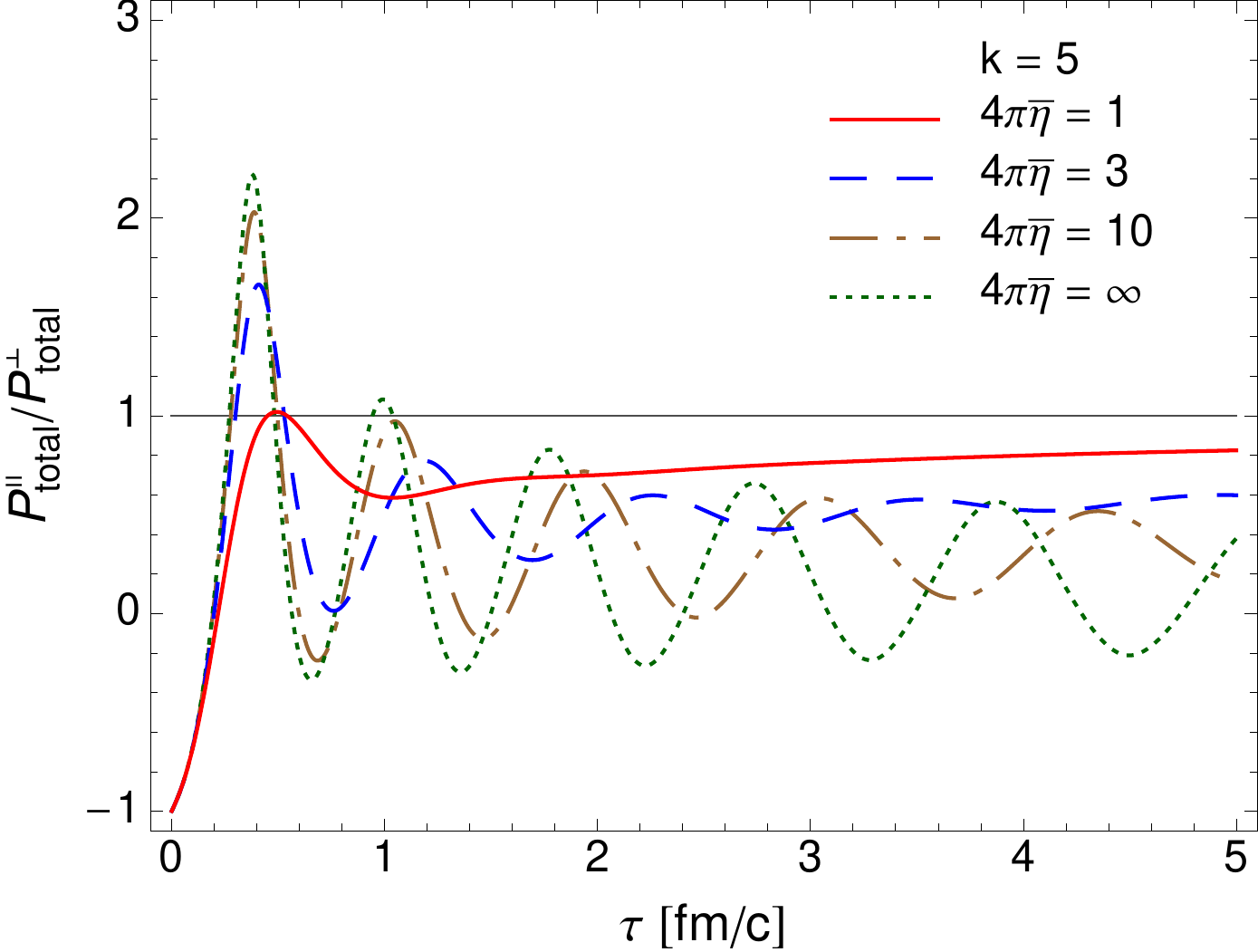}
\quad
\includegraphics*[width=6.cm]{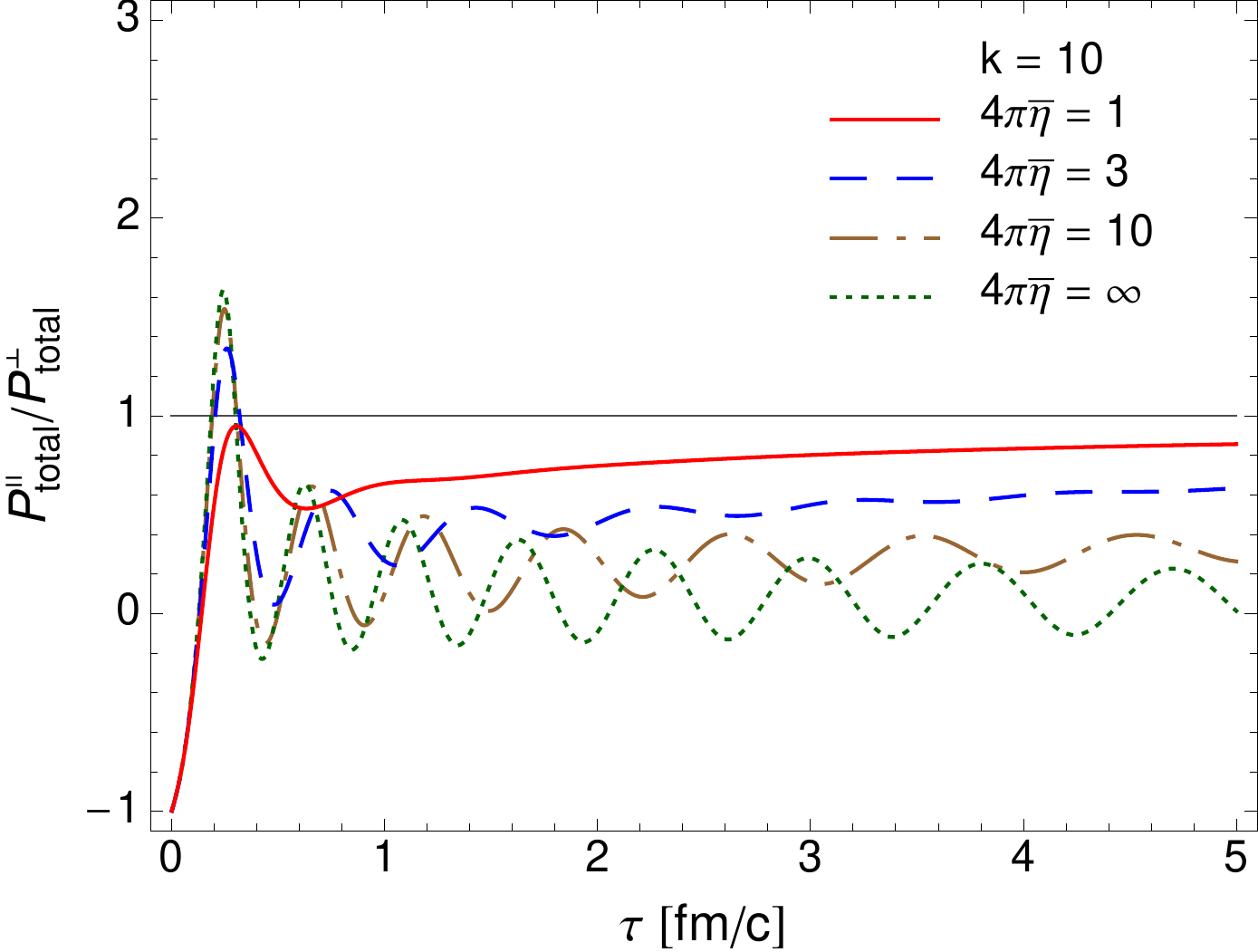}
\caption{(Color online) Time dependence of the ratio $P^{\rm total}_\parallel/P^{\rm total}_\perp$ for $k=5$ (left) and $k=10$ (right), and for different values of the viscosity. }
\label{fig:pres}
\end{center}
\end{figure}

In the context of thermalization, it is important to study different pressure components of the system. The ratio of the total (particles+field) longitudinal pressure and the total (particles+field) transverse pressure is shown in Fig.~\ref{fig:pres}. Since the longitudinal chromoelectric field gives a negative contribution to the longitudinal pressure and a positive contribution to the transverse pressure, the main effect of the field is to lower the $P_\parallel(\tau)/P_\perp(\tau)$ ratio calculated for the system of particles only. This effect is the strongest at the beginning of the production process when the color fields are the strongest. For later times and the collissionless case, $4\pi{\bar \eta}=\infty$, the ratio of the two pressures strongly oscillates and its average value is significantly smaller than 1. However, with decreasing viscosity, the ratio of the two pressures gets closer to unity, which reflects tendency of the system to reach local thermal equilibrium. Qualitatively this behavior is similar to that found in \cite{Gelis:2013rba,Berges:2013fga}.

\section{Conclusions}

We have studied thermalization of the quark-gluon plasma produced by decays of color flux tubes possibly created at the early stages of relativistic heavy-ion collisions. A novel feature of our approach is the implementation of the viscosity of the produced quark-gluon plasma in terms of a constant ratio of the shear viscosity coefficient to the entropy density, ${\bar \eta} = \eta/\sigma$ = const. For the lowest (KSS) value of the ratio of the shear viscosity to the entropy density, $4\pi{\bar \eta} = 1$, the analyzed system approaches the viscous-hydrodynamics regime within  1--2 fm/c. On the other hand, for larger values of the viscosity, $4\pi{\bar \eta} \geq 3$, the collisions in the plasma are not efficient to destroy collective phenomena in the plasma, which manifest themselves as oscillations of different plasma parameters.

{\bf Acknowledgments:} R.R. was supported in part by the Polish National Science Center grant with decision No. DEC-2012/07/D/ST2/02125. W.F. was supported in part by the Polish National Science Center grant with decision no. DEC-2012/06/A/ST2/00390.

\end{document}